\documentclass{article}
\usepackage{graphicx} 
\usepackage[table]{xcolor}
\usepackage{color, colortbl}

\usepackage{authblk}
\providecommand{\keywords}[1]{\textbf{\textit{Keywords:}} #1}

\title{Will Central Bank Digital Currencies (CBDC) and Blockchain Cryptocurrencies Coexist in the Post Quantum Era? } 
\author[1]{Abraham Itzhak Weinberg}
\author[2]{Pythagoras Petratos,}
\author[3]{Alessio Faccia }
\affil[1]{AI-WEINBERG, AI Experts, Tel Aviv, Israel, aviw2010@gmail.com}
\affil[2]{EFA, Coventry University, UK, p.pythagoras@yahoo.com}
\affil[3]{University of Birmingham Dubai, Dubai, UAE, alessio.faccia@gmail.com}

\begin{document}
\maketitle
\begin{abstract}
This paper explores the coexistence possibilities of Central Bank Digital Currencies (CBDCs) and blockchain-based cryptocurrencies within a post-quantum computing landscape. It examines the implications of emerging quantum algorithms and cryptographic techniques such as Multi-Party Computation (MPC) and Oblivious Transfer (OT). While exploring how CBDCs and cryptocurrencies might integrate defenses like post-quantum cryptography, it highlights the substantial hurdles in transitioning legacy systems and fostering widespread adoption of new standards.
The paper includes comprehensive evaluations of CBDCs in a quantum context. It also features comparisons to alternative cryptocurrency models. Additionally, the paper provides insightful analyses of pertinent quantum methodologies. Examinations of interfaces between these methods and blockchain architectures are also included. The paper carries out considered appraisals of quantum threats and their relevance for cryptocurrency schemes. Furthermore, it features discussions of the influence of anticipated advances in quantum computing on algorithms and their applications.
The paper renders the judicious conclusion that long-term coexistence is viable provided challenges are constructively addressed through ongoing collaborative efforts to validate solutions and guide evolving policies.
\end{abstract}

\keywords{Quantum Money, Central Bank Digital Currencies (CBDC),  Decentralized Finance (DeFi), Blockchins, Cryptocurrency, Oblivious Transfer/ Protocol (OT), Multi-Party Computation (MPC)}

\section{Introduction}
When considering CDBC and blockchain-based cryptocurrencies, the famous lines from the poem 'The Road Not Taken' by Robert Frost come to mind: 'Two roads diverged in a yellow wood, And sorry I could not travel both and be one traveler, long I stood' \cite{cockburn2020road}. However, unlike the poem, in this paper, we will discuss the possibility of choosing both roads in the post-quantum era. Metaphorically, we can liken the practical aspects of this issue to those in quantum physics, known as "quantum superposition", where elementary particles exist in two parallel states simultaneously until their location is observed at a specific point \cite{lewis2020unification}. The quantum world supports both digital currencies approaches, and it is up to the decision-maker to determine what to choose. \\

In this era, a notable trend is our significant progress towards the post-quantum era. This phase denotes the time following the advancement of quantum computing to a level where it can potentially compromise prevalent public-key cryptosystems like Rivest–Shamir–Adleman (RSA) and Elliptic Curve Cryptography (ECC) \cite{roy2019survey}.
In parallel, we see a rise in popularity of CDBC \cite{auer2020rise} side by side to rise in usage of cryptocurrency \cite{sahoo2017bitcoin}.
Within this context, the importance of protocols such as Oblivious Transfer/Protocol (OT) and Multi-Party Computation (MPC) remains paramount in ensuring secure and private transactions \cite{yadav2022survey}.
OT is a cryptographic protocol where a sender transmits one of potentially multiple pieces of information to a receiver without knowing which specific piece the receiver obtains, enabling the receiver to access information from the sender without disclosing the chosen item's identity \cite{yadav2022survey}. The sender merely acknowledges the transfer but remains unaware of the item's index. OT finds utility in scenarios such as privacy-preserving data mining and secure two-party computation. Notable examples encompass 1-out-of-2 OT, where the sender presents two data pieces for the receiver to select from discretely, and 1-out-of-N OT, where the sender offers N data pieces for selection without revealing the specific item index \cite{amiri2021imperfect}. OT serves as a foundational element in secure two-party and multi-party computations, facilitating collaborative function computations over private inputs while maintaining confidentiality.\\
MPC extends privacy protection to scenarios involving multiple entities collaborating to compute functions over their respective confidential inputs \cite{evans2018pragmatic}. For instance, banks can collectively calculate credit scores without disclosing individual customer data. In MPC, input data is partitioned and shared among distinct, non-colluding parties to prevent any single entity from accessing another's confidential information. Subsequently, algorithms operate on these distributed data shares to produce outputs in a privacy-preserving manner. MPC methodologies commonly rely on secret sharing, garbled circuits, and cryptographic tools like Homomorphic Encryption (HE)\footnote{A cryptographic method enabling calculations on encrypted data without the need for decryption \cite{acar2018survey}.} and Zero-Knowledge Proofs (ZKPs) \footnote{A protocol where one party can prove the truth of a given statement to another party without revealing any additional information beyond the statement's truth \cite{morais2019survey}.}, with OT serving as a fundamental component in many MPC protocols. The application spectrum of MPC spans privacy-centric technologies, blockchain ecosystems, and distributed analytics involving sensitive data, offering a secure framework for collaborative computation while safeguarding individual data privacy.

\subsection{Reasons for Rise of Central Bank Digital Currencies (CDBC)}
Based on a recent survey, at the conclusion of 2021, 90\% of the 81 respondent central banks in Europe were actively exploring the potential for a CBDC \cite{kosse2022gaining}. In the latter half of 2023, the traction supporting CBDCs has persisted. Recent findings from CBDC tracker reveal that 130 nations are currently delving into CBDCs, encompassing 98 percent of global GDP \cite{Chhangani2023CBDC}.\\
The surge in the development of CBDCs can be attributed to several key factors \cite{nanez2020reasons}. Firstly, there is an escalating demand for digital payments as online commerce expands and societies transition towards cashless transactions, necessitating government-backed digital currencies to accommodate modern payment methods, a need that CBDCs fulfill. Additionally, the emergence of decentralized cryptocurrencies like Bitcoin has posed a challenge to traditional currency monopolies, prompting central banks to explore hybrid digital currencies through CBDCs to maintain control while adapting to the changing landscape.\\
Moreover, CBDCs hold promise in promoting financial inclusion by offering a universally accessible digital currency option, thereby expanding banking services to the unbanked and underbanked populations \cite{sakharov2021central}. Furthermore, the utilization of digital currencies supported by distributed ledgers could revolutionize payment systems by enabling real-time, cost-effective domestic and cross-border transactions compared to conventional financial infrastructure. CBDCs also present an opportunity for central banks to introduce additional policy tools, allowing for tailored money issuance or transaction incentives to inject economic stimulus directly into the economy during periods of downturn. \\
Many central banks view CBDCs as a means of modernizing their payment systems and embracing blockchain-based technologies that drive future financial innovations. Lastly, the development of CBDCs serves as a preemptive strategy for central banks to proactively address potential risks such as the growing dominance of non-sovereign cryptocurrencies or the disruptive influence of emerging technologies like quantum computing on existing systems. In essence, the rise of CBDCs is driven by the increasing demand for digital payments, competition from cryptocurrencies, financial inclusion objectives, central banks' pursuit of innovative monetary policy tools, and the push for technological advancement in payment systems.

\subsection{Reasons for Rise in Usage of Crypto-assets}
Preserving privacy in CBDCs poses a challenge. The core priorities of transparency, anonymity, traceability and compliance posed by a CBDC model introduce tensions with privacy that are challenging to resolve through technology and policy \cite{bashir2016motivates}. It will require novel solutions. This is one of the reasons that support the usage of cryptocurrencies.\\
The rise in crypto-assets \footnote{A crypto-asset is a digital representation of value or entitlement that can be electronically transferred or stored through distributed ledger technology or comparable mechanisms \cite{parrondo2023cryptoassets}.} as well as cryptocurrencies' popularity can be attributed to several key reasons \cite{sahoo2017bitcoin}. Firstly, it serves as an alternative store of value to fiat currencies, appealing to individuals seeking a hedge against inflation and currency devaluation due to its capped supply, making it an attractive long-term investment. Moreover, cryptocurrency's price volatility has drawn in speculators aiming to capitalize on its increasing valuation over time, attracting a broader investor base. Its utility for international payments offers a quick and cost-effective way to transfer value across borders independently of traditional banking systems, enhancing its use for remittances and global commerce. Additionally, cryptocurrency provides a level of financial anonymity unparalleled in conventional banking systems, enhancing its attractiveness to those valuing privacy and pseudonymity in transactions \cite{gulled2018bitcoins}. Ideologically, cryptocurrency aligns with the desire for decentralization and independence from governmental or corporate influence, appealing to individuals seeking a truly decentralized currency. The growing acceptance of cryptocurrencies by major payment processors, merchants, exchanges, and the proliferation of cryptocurrency-supporting ATMs have widened its accessibility and usability, expanding its user base. Functioning outside traditional financial systems, cryptocurrency offers an alternative financial network that is immune to censorship, appealing to those interested in utilizing censorship-resistant applications and money. The increasing interest in blockchain technology by businesses has fueled general awareness and excitement around cryptocurrencies like Bitcoin, further boosting its popularity. Ultimately, cryptocurrency's scarcity, global accessibility, privacy features, and decentralized nature drive both speculative investment and practical adoption, establishing it as a versatile asset with a range of valuable use cases.

\subsection{Development of Quantum Technology}
As of now, we are not yet in the post-quantum era, a period that will dawn once quantum computers attain the capability to break mainstream public-key cryptography such as RSA and ECC \cite{awschalom2021development}. While the largest quantum computers currently boast around 100 qubits, the task of breaking AES-256 or factoring a 2048-bit RSA key demands millions of qubits, a milestone that remains distant \cite{bavdekar2022post}.\\
Quantum computing experts project that realizing quantum computers capable of breaching current cryptographic standards is still a decade or two away at the earliest, given the ongoing progress in the field. Presently, we find ourselves in the 'noisy intermediate-scale quantum' era, characterized by quantum computers with limited qubits and high error rates. These machines are primarily utilized for small proof-of-concept demonstrations rather than general computational tasks \cite{vidick2023introduction}. In addition, we find several recent trials for attacking to military grade encryption using quantum computer that seems ambitious and advance the post quantum era \cite{chinS2024swayne,milg2024buchanan}. \\
Going back to 1983, Stephen Wiesner's paper introduced groundbreaking ideas well ahead of their time. In it, he proposed using delicate quantum states as unforgeable banknotes, drawing on emerging studies of quantum physics \cite{wiesner1983conjugate,lee2020future}.
Specifically, Wiesner envisioned generating quantum money consisting of non-orthogonal quantum states, such as polarization states of single photons. These states could not be perfectly distinguished or cloned due to the fundamental no-cloning theorem of quantum mechanics.\\
As such, any attempt to counterfeit a quantum banknote by copying or measuring it would necessarily introduce errors. The genuine banknote could then be verified by carefully measuring in designated bases. This exploiting of intrinsic quantum properties for authentication marked the earliest concept of quantum currency.\\
Wiesner's work was also seminal in that it helped lay the theoretical foundation for later advances in quantum information science and cryptography \cite{wiesner1983conjugate}. His money proposal directly inspired the seminal ideas behind quantum key distribution, wherein non-orthogonal states allow securely encoding private keys \cite{amiri2017quantum}.\\
While initially a theoretical thought experiment, Wiesner's work kicked off serious research efforts to devise practical quantum money schemes. It demonstrated how quantum physics could enable fundamentally new approaches to issues like counterfeiting prevention - applications that are only now becoming technologically feasible decades later with improvements in quantum control and devices.\\
Overall, Wiesner's paper pioneered the study of using quantum effects for tasks like encryption, key distribution and monetary systems. It served as a prescient early thought experiment that remains highly influential on modern developments in quantum technologies and their information-theoretic applications.\\
Blockchain technology and cryptocurrencies, starting with Bitcoin in 2009, have revolutionized digital finance and payments \cite{hashemi2020cryptocurrency}. By leveraging distributed ledger systems secured through cryptography, blockchains enable decentralized networks for value exchange without reliance on centralized intermediaries like banks.\\
This ushered in new possibilities for trustless, permissionless value transfer across borders via digital assets. It appealed strongly to a vision of open, democratic money for the internet age. Crypto also spurred innovation through technologies like smart contracts and decentralized applications \cite{de2017survey}.\\
Within just over a decade, blockchain networks and cryptocurrencies grew exponentially in users, applications and total value. Bitcoin in particular saw rising adoption as an alternative currency and speculative asset \cite{hairudin2022cryptocurrencies}. This rapid rise disrupted traditional finance and forced rethinking of money and its underlying technologies.
However, most blockchains rely on asymmetric key cryptography like RSA and ECC to digitally sign transactions and protect wallet addresses/keys \cite{chalkias2018blockchained}. While sufficiently secure today, these systems will be vulnerable once quantum computers achieve substantial processing power.\\
Specifically, algorithms like Shor's could break the prime number factorization and discrete logarithm problems that underpin RSA and ECC \cite{shor2012quantum}. This threatens the core security of blockchains by enabling forgery of signatures and theft of funds from public keys. Urgent upgrades are needed to ward off these quantum hacks.\\
The looming quantum computing threats thus challenge the long-term viability of current blockchain systems and stored crypto value, unless they pivot to quantum-resistant cryptographic schemes still being developed. This context introduces risk and uncertainty to the technology.
Blockchain technology and cryptocurrencies have introduced revolutionary changes to digital financial system in recent years \cite{de2017survey}. The ability to conduct secure peer-to-peer value transfers without centralized intermediaries has significant implications. Blockchains provide an open, decentralized platform for developing new applications of digital assets, smart contracts, and other FinTech innovations. Cryptocurrencies like Bitcoin have also seen rising adoption as alternative payment networks and speculative assets.\\
However, most current blockchain networks rely on encryption algorithms that are susceptible to being broken by quantum computers. In particular, public-key cryptography schemes like RSA and ECC form the backbone of how transactions, wallets, and nodes are secured on blockchains. These asymmetrical encryption algorithms work by factoring large prime numbers, but Peter Shor brilliantly devised an algorithm in 1994 that can solve this problem exponentially faster on a sufficiently powerful quantum computer \cite{shor1994algorithms}.\\
Shor's algorithm threatens to render the encryption underpinning blockchains totally insecure \cite{shor2012quantum}. A quantum computer able to run Shor's algorithm would be able to derive the private keys needed to spend cryptocurrencies from public keys, impersonate nodes on a blockchain network, and rewrite transaction histories. This applies to the signature schemes, address generation, and other core cryptographic primitives used across blockchains today from Bitcoin to Ethereum.\\
While quantum computers with this capability may still be 10-20 years away, the long-term implications are serious. Blockchain networks, DeFi applications, and the value and integrity of held cryptocurrencies could all be jeopardized without modifications to use quantum-resistant cryptography \cite{kearney2021vulnerability}. This highlights the urgent need for the blockchain industry to start transitioning to post-quantum secure algorithms and implementations.\\
In response to these quantum computing threats and the rise of private cryptocurrencies, central banks have begun exploring the development of their own state-backed digital currencies. Known as CBDCs, these are digital forms of fiat currency issued and governed by a country's monetary authority \cite{yang2022decentralization}.
Most CBDC prototypes and pilot programs utilize a centralized database or distributed ledger architecture rather than a public blockchain \cite{sethaput2023blockchain}. This model gives central banks full control over key functions like money issuance, transaction management, interest rates, and compliance. Proponents argue CBDCs can provide many of the benefits of digital cash while preserving national monetary sovereignty.\\
Some key benefits central banks hope to achieve with CBDCs include increasing payment efficiency, reducing transaction costs, expanding financial inclusion, and facilitating fiscal stimulus measures \cite{shkliar2020phenomenon}. It also gives them a secure alternative to private cryptocurrencies at a time when quantum computers may undermine current blockchain networks. Countries like China are already testing CBDC projects aimed at replacing cash usage \cite{allen2022fintech}.
Meanwhile, the blockchain industry is proactively developing solutions to secure digital assets against quantum attacks. This involves transitioning blockchain networks, addresses, and cryptography to post-quantum secure algorithms theorized to withstand even fault-tolerant quantum computers \cite{allende2023quantum}. Standards are being proposed for quantum-resistant signature schemes, zero-knowledge proofs, and other primitives.\\
Overall, both central banks and the blockchain industry see the need to evolve for the quantum era. CBDCs aim to provide quantum-secure centralized digital fiat money. And continuous work on post-quantum cryptography seeks to future-proof decentralized assets on blockchains and allow them to coexist alongside CBDCs.
As the capabilities of quantum computers continue advancing, it raises the crucial question of how digital currencies might evolve in response. Will there be space for both centralized CBDCs controlled by central banks as well as decentralized cryptocurrencies powered by blockchain technology to coexist? Or will one dominate the other as quantum computing disruptions play out?\\
There are reasonable arguments on both sides. On one hand, central banks may prefer the sovereignty and oversight afforded by CBDCs, pushing for them to become the dominant digital legal tender. Powerful governments could potentially restrict or ban private cryptocurrencies in such a scenario.
However, blockchain networks are also proactively working to bolster security and make cryptocurrencies quantum-resistant. Demand for decentralized, censorship-resistant assets like Bitcoin remains strong among many users as well. This suggests cryptocurrencies may find ongoing utility and market demand even in a post-quantum world.\\
Some possibilities for coexistence could emerge if central banks integrate certain blockchain or distributed ledger features into CBDCs to gain efficiencies, while maintaining centralized issuance controls. Hybrid public-private models may also see central bank-backed digital tokens running on permissioned blockchain platforms.
Overall, how dominant each model becomes will likely depend on evolving technical standards as well as policy priorities around financial sovereignty, inclusion, and innovation. New collaborative frameworks between government monetary authorities and the cryptocurrency industry may also need to be explored to tap their respective advantages for users.

\section{A Central Bank Digital Currency (CBDC) in Post Quantum Era}
CBDC is a digital form of currency issued by a central bank that can be utilized by the general public \cite{ahmed2022roles}. It serves as the digital counterpart of physical cash, providing a convenient and secure means of conducting transactions.
Distinguished from physical cash, a CBDC solely exists in digital form and represents a direct liability of the central bank on its balance sheet. This means that individuals and businesses can hold CBDC accounts directly with the central bank, enhancing accessibility and eliminating the need for intermediaries.
CBDCs are primarily designed to ensure financial stability and uphold sovereignty over monetary policy. By offering citizens continued access to central bank money, even as the use of physical cash diminishes, CBDCs contribute to the preservation of financial systems. \\
One potential advantage of CBDCs is their ability to expedite domestic and international fund transfers while reducing transaction costs compared to physical cash and credit cards \cite{lee2021global}. Moreover, CBDCs hold the promise of expanding financial inclusion by granting broader access to financial services.
Nevertheless, the widespread implementation of CBDCs also entails certain risks. Technical failures could cause disruptions, and concerns arise regarding the privacy and anonymity of transactions, as well as the potential for increased surveillance powers of governments and central banks over individuals' financial activities \cite{mu2022cbdc}. Cybersecurity poses a significant technical challenge that must be addressed comprehensively.\\
Many central banks, including the Bank of England, the European Central Bank, China's central bank, and the Federal Reserve, are actively researching or piloting CBDCs \cite{auer2020rise}. However, it is important to note that most projects are still in the research or pilot phase, indicating that comprehensive implementation is still in progress.
The extensive adoption of CBDCs could potentially disrupt the existing structure of the banking system and private payment systems such as credit cards. Thus, proper regulations and frameworks would need to be established to govern their usage effectively and maintain financial stability in the evolving landscape of digital currencies.

\section{Central Bank Digital Currency (CBDC) versus Cryptocurrencies and Blockchains}
When comparing CBDC to cryptocurrencies and blockchains, several key distinctions arise as can be seen in Table \ref{tab:cbdc-vs-cryptocurrencies}.
Firstly, CBDCs are issued and backed by a central bank, representing the authority of a particular country, whereas cryptocurrencies operate in a decentralized manner and are not issued by any central authority \cite{zhang2022blockchain}.
In terms of use cases, CBDCs are primarily intended as legal tender for retail transactions, aiming to replace physical cash. On the other hand, cryptocurrencies are often used as speculative assets or for facilitating borderless payments.
Stability is another differentiating factor. CBDCs are designed as sovereign currencies with the objective of maintaining price stability. In contrast, cryptocurrencies are known for their volatility and lack of stability.\\
Privacy features also differ between CBDCs and cryptocurrencies. While CBDCs may incorporate higher privacy standards compared to physical cash, they typically provide less privacy than many cryptocurrencies, which often strive for anonymity. Central banks often seek oversight capabilities and regulatory compliance.
Technologically, CBDCs are likely to utilize permissioned distributed ledgers tailored to the specific requirements of central bank operations. This stands in contrast to the permissionless blockchains commonly associated with cryptocurrencies.
Accessibility is another aspect to consider. CBDCs are designed for widespread retail use by citizens, functioning as a digital representation of cash. Cryptocurrencies, on the other hand, tend to cater more to early adopters and tech-savvy individuals.\\
Regulation plays a significant role as well. CBDCs can enforce strong compliance measures and adhere to anti-money laundering rules set by governments and regulators. In contrast, cryptocurrencies currently operate with minimal regulation.
Finally, when it comes to adoption, CBDCs aim for universal domestic usage within their respective countries, while cryptocurrencies still face challenges in achieving mass-market diffusion and widespread acceptance by merchants.
In summary, CBDCs prioritize stability, control, and oversight over innovation, while offering familiar attributes of sovereign fiat currencies in a digital format.

\begin{table}
\centering
\begin{tabular}{|p{2cm}|p{5.8cm}|p{5.8cm}|}
\hline
\rowcolor{blue!30}
\textbf{Aspect} & \textbf{Central Bank Digital Currencies (CBDC)} & \textbf{Cryptocurrencies/Blockchains} \\
\hline
\hline
Issuer & Central bank & Decentralized \\
\hline
Use case & Retail transactions, replacing cash & Speculative assets, borderless payments \\
\hline
Stability & Designed for price stability & Volatile \\
\hline
Privacy & Higher than physical cash, but less than most cryptocurrencies & Aim for anonymity \\
\hline
Technology & Permissioned distributed ledgers suitable for central bank operations & Permissionless blockchains \\
\hline
Accessibility & Widespread retail use & Early adopters, tech-savvy users \\
\hline
Regulation & Strong compliance measures, regulation enforced & Mostly unregulated \\
\hline
Adoption & Universal domestic usage & Lack of mass-market diffusion, limited merchant acceptance \\
\hline
\end{tabular}
\caption{Comparison between Central Bank Digital Currencies (CBDC) and Cryptocurrencies/Blockchains.}
\label{tab:cbdc-vs-cryptocurrencies}
\end{table}
\subsection{Comparative Analysis}
Comparing CBDCs and blockchain-based cryptocurrencies in a post-quantum computing landscape involves assessing their abilities to withstand quantum computer threats and integrating quantum-safe security mechanisms effectively \cite{de2024applying} as can be seen in Table \ref{tab:cbdc-vs-crypto}.\\
CBDCs, being centralized systems under governmental and central bank jurisdiction, could swiftly adopt post-quantum cryptographic standards once they mature, leveraging software updates for algorithm transitions. Nonetheless, updating existing CBDCs to quantum-safe versions may present technical obstacles, necessitating the replacement of all previously issued digital currency. On the other hand, blockchain-based cryptocurrencies function on decentralized networks, requiring coordinated consensus across all nodes for cryptography upgrades. The decentralized nature may prolong the transition to post-quantum algorithms due to the need for broad community agreement.\\
Nevertheless, cryptocurrencies could exhibit inherent resilience as transactions are typically pseudonymous, reducing quantum computers' impact on privacy and fungibility. Both approaches could incorporate strategies like MPC and HE to safeguard transaction privacy against quantum threats. However, implementing these techniques in large-scale payment networks poses performance challenges. While CBDCs possess resources and centralized control for adopting new technologies, cryptocurrencies have shown greater innovation in enhancing blockchain efficiency.

\begin{table}
\centering
\begin{tabular}{|p{2cm}|p{5.8cm}|p{5.8cm}|}
\hline
\rowcolor{blue!30}
& \textbf{Central Bank Digital Currency (CBDC)} & \textbf{Blockchain-based Cryptocurrency} \\
\hline
\hline
Ability to adopt post-quantum cryptography & Potential advantage - centralized systems can implement new algorithms through software updates. & Upgrading cryptography requires consensus across decentralized networks, making transition longer. \\
\hline
Resilience against quantum computers & Transitioning existing systems to quantum-safe versions poses technical challenges and requires replacing all currency. & Some cryptocurrencies may be inherently more resistant due to pseudonymous transactions limiting impact on privacy. \\
\hline
Integration of quantum-safe techniques like MPC/HE & Could apply new techniques like MPC and HE but brings significant performance hurdles to large-scale payment networks. & Also possible to integrate new techniques, but introduces bigger technical challenges to blockchain scaling. \\
\hline
Governance over cryptographic modernization & Strong centralized governance allows for more control over updates. & Upgrades require coordination across many parties in decentralized ecosystem. \\
\hline
\end{tabular}
\caption{Comparative Analysis of Central Bank Digital Currency (CBDC) and Blockchain-based Cryptocurrencies in a Post-Quantum Era}
\label{tab:cbdc-vs-crypto}
\end{table}

\section{Connection between Oblivious Transfer, Multi-Party Computation (MPC) and Blockchain}
Oblivious Transfer (OT) and Multi-Party Computation (MPC) play essential roles in enhancing privacy within blockchain systems \cite{zhong2020secure}.
While blockchain provides a decentralized trust infrastructure, on-chain computations often lack privacy. However, OT and MPC offer solutions to incorporate privacy into blockchain applications.
OT and MPC enable decentralized protocols to evaluate functions and perform transactions over private inputs from multiple participants, preserving privacy. This opens the door to valuable use cases such as private smart contracts, and decentralized finance.
Within blockchain, OT finds applications in private coin shuffling/mixing, creating anonymity sets for wallet addresses, mixing outputs of coinjoins, and facilitating private coin issuance and transfers.
MPC empowers decentralized exchanges and oracles to compute order books, process trades, and query off-chain data while safeguarding the privacy of user inputs and outputs.
By utilizing MPC and OT to generate zero-knowledge proofs, blockchain transactions and computations can be validated without revealing unnecessary private details \cite{zhou2021using}.
Decentralized voting protocols and identity systems leverage OT and MPC to protect user privacy during interactions within the blockchain.
Research is actively exploring the integration of efficient OT and MPC with blockchain consensus mechanisms. This integration aims to enable trustless and permissionless execution of private smart contracts and decentralized Applications (dApps).
Standardizing secure blockchain primitives such as OT, MPC, and Zero-Knowledge Succinct Non-Interactive Argument of Knowledge (zk-SNARKs) \footnote{A zk-SNARK is a protocol that enables a prover to demonstrate to a verifier the truth of a statement regarding confidential data without disclosing the data itself.} is an ongoing endeavor \cite{chen2022review}. This standardization work aims to facilitate the development of privacy-enhanced decentralized applications on a larger scale.

\subsection{Multi-Party Computation (MPC) and its impact on Blockchain and Cryptocurrencies}
MPC is a remarkable technique that facilitates collaborative computation among two or more parties, ensuring the privacy of their inputs. Through MPC, these parties can jointly execute a function while only obtaining the output and remaining oblivious to each other's specific inputs \cite{halevi2019fully}.\\
Rapid progress in developing quantum computers poses significant risks to the security of blockchain networks and digital payment systems, as quantum algorithms may threaten the cryptographic foundations that underpin these technologies. As a means to bolster defenses against future quantum computer attacks, integrating secure MPC protocols into blockchains has emerged as a promising mitigation strategy. MPC enables distributed computations on private inputs without revealing intermediate values, relying on secret sharing schemes believed to withstand large-scale quantum computers. \\ 
One of the key advantages of MPC is its ability to enable secure distributed computing, even in scenarios where some participants may be untrusted or tempted to cheat. As long as at least one party acts honestly, both privacy and correctness are preserved.
MPC leverages cryptographic methods and secure computation techniques such as secret sharing to evaluate functions on private data securely. Inputs are divided into shares and distributed among the parties, ensuring that no individual party gains access to the complete input.\\
The applications of MPC span a wide range of domains, including privacy-preserving data analytics, blockchain voting, and financial services. It empowers computations on sensitive information like medical records or tax files without compromising the confidentiality of the underlying data.
However, MPC does come with its own set of challenges \cite{harbi2023model}. Performance can be a concern since secure computations tend to be slower than regular computations. Scalability to handle large inputs and outputs is another area of focus, as well as efficiently incorporating new functions into the MPC framework.
Recent research has made significant strides in improving the efficiency of MPC and expanding its capabilities to accommodate more complex functions. Industries are increasingly embracing MPC for collaborative analytics and other use cases where privacy is of paramount importance.\\
Notable companies engaged in MPC include Anthropic, Duality Technologies, and Partisia Blockchain, among others. Furthermore, efforts are underway to standardize MPC protocols in order to foster widespread adoption and application in real-world scenarios.

\subsubsection{Potential of MPC and Post-Quantum Security}
The potential of combining secure MPC with post-quantum security measures presents a layered security approach where MPC operates at the application layer alongside post-quantum cryptography at the algorithmic layer, offering defenses against quantum attacks at multiple levels \cite{khan2024future}. \\
Through distributed computations and post-quantum secure secret sharing within an MPC framework, a robust defense strategy can be achieved, ensuring protection even as individual techniques progress towards maturity. This combined approach enhances resilience by reducing single points of failure, a strength not achievable when relying solely on either method \cite{feneuil2023post}.  \\
By integrating quantum-safe techniques within MPC protocols, systems can be "future-proofed" against evolving quantum computing capabilities. Additionally, synergies between MPC and post-quantum foundations, such as lattice-based assumptions, can enhance certain MPC protocols like multiparty threshold cryptography. Layering MPC over post-quantum cryptography maximizes defenses, enabling systems to guard against both immediate and long-term quantum threats effectively. This area of research, directly integrating quantum-safe MPC into applications like blockchains, remains relatively unexplored, presenting opportunities for novel contributions. In essence, the potential of combining MPC with post-quantum security offers a comprehensive and forward-looking security approach that addresses short and long-term quantum computing risks, warranting further investigation and analysis in this underexplored domain.

\subsubsection{Oblivious Protocol/ Transfer (OT)}
Oblivious Transfer (OT) plays a crucial role in both secure multiparty computation and cryptography, serving as a fundamental building block \cite{burra2021high}. Following are some key aspects of Oblivious Transfer protocols.
OT allows a sender to transmit one out of several pieces of information to a receiver without revealing which specific item was sent, while ensuring that the receiver only learns the requested item. This is achieved by having the sender encrypt all possible items and send them to the receiver, who can then privately decode the item they desire without disclosing their choice to the sender.\\
The 1-out-of-2 OT (OT12) protocol is the fundamental component of Oblivious Transfer \cite{amiri2021imperfect}. It enables the receiver to obtain one of two items from the sender without revealing to the sender which item was received. This forms the basis for more complex OT constructions.
OT extensions enhance the efficiency of OT by allowing a large number of OTs to be performed with sub-linear communication complexity. This is accomplished by initially running a small number of OT12 protocols.
The significance of OT in secure computation lies in its ability to enable parties to privately select their inputs for a computation, safeguarding the confidentiality of their inputs from other participants.
The applications of OT are diverse and far-reaching. They encompass private database access, electronic voting systems, secure cryptocurrency transactions, and secret sharing protocols, among others.
Implementations of OT rely on various techniques such as Yao's garbled circuits, HE, or assumptions of semi-honest behavior by the involved parties \cite{saleem2018recent}.\\
Efficiency and widespread usability of OT remain active areas of research within multiparty computation (MPC), as researchers strive to enhance the performance and practicality of OT protocols for real-world applications \cite{huang2011efficient}.

\section{Quantum  Oblivious Transfer (OT) and Multi-Party Computation (MPC) and Their Influence on Blockchains and Cryptocurrencies}
Quantum Oblivious Transfer (QOT) and Quantum Multi-party Computation (QMPC) offer exciting prospects for secure computation and privacy in distributed systems \cite{lemus2020generation,sarkar2024efficient}. These protocols utilize principles of quantum information and entanglement to enable computations on private inputs from multiple parties, potentially providing even stronger privacy guarantees than classical MPC.
In quantum OT, the sender encodes inputs using non-orthogonal quantum states that the receiver can distinguish but cannot perfectly clone or copy due to the fundamental laws of quantum mechanics \cite{santos2022quantum}. Protocols based on quantum teleportation or dense coding have been developed to realize quantum OT, although practical implementations present challenges that need to be addressed. \\
Quantum MPC allows for the secure evaluation of quantum or classical functions using private quantum inputs. This has applications in areas such as private database queries and financial computations. Achieving secure quantum OT and MPC in the presence of dishonest parties requires building blocks like quantum bit commitment and coin flipping. 
Currently, quantum OT and MPC are active areas of research, focusing on designing efficient protocols, overcoming noise and errors, and implementing them using quantum technologies such as trapped ions or photons. Once these challenges are overcome, they could pave the way for entirely new forms of multiparty applications, including blind quantum computation, which is impossible to achieve using classical methods. \\
Significant theoretical progress has been made in quantum OT and MPC, but there are still hurdles to address. Scalability, noise, and integration with classical distributed systems are among the challenges that need to be overcome for practical implementation.
In summary, quantum OT and MPC protocols leverage the principles of quantum mechanics to provide stronger privacy guarantees in distributed computing. However, it is crucial to address the technical obstacles and challenges associated with these protocols to fully realize their potential in enhancing privacy and security in distributed systems.\\
Quantum OT and MPC have the potential to significantly impact blockchains and cryptocurrencies, introducing new possibilities and challenges:
One area of impact is privacy-enhanced transactions. By leveraging quantum OT and MPC, blockchain transactions can be conducted with full privacy. This means computations can be securely performed on private inputs, such as wallet balances, without revealing transaction amounts on the public ledger.
The use of quantum OT in cryptocurrencies can lead to the development of anonymous digital assets. Similar to CoinJoin on traditional blockchains, quantum OT can facilitate private coin mixing or shuffling, resulting in cryptocurrencies with stronger anonymity sets.
Quantum MPC enables the execution of smart contracts with private inputs and outputs. This opens up opportunities for private Decentralized Finance (DeFi) applications, including private loans, betting markets, auctions, and more.\\
To maintain security over long time scales, the implementation of quantum-safe cryptosystems and protocols may require specialized quantum hardware for key generation and storage. Trusted hardware becomes crucial in ensuring the security of quantum-resistant systems.
In terms of consensus mechanisms, post-quantum solutions need to incorporate digital signatures, commitments, and other quantum-safe elements to resist both classical and quantum computers \cite{ciulei2022preparation}.
Compliance challenges arise with the introduction of quantum protocols. Traceability and monitoring requirements for Anti Money Laundering (AML) and Know Your Customer (KYC) regulations become more complex compared to transparent blockchains \cite{thommandru2023recalibrating,van2023cryptocurrency}.
The design of cryptocurrencies can be upgraded with concepts from quantum money, such as the authentication of quantum coin states. These concepts can influence the design of digital assets issued on permissioned quantum-resistant ledgers. \\
However, there are research challenges that need to be addressed. Significant progress is still required to scale quantum computation for real-world systems and to effectively interface quantum technologies with existing blockchain architecture and decentralized networks.
In the long term, the practical implementation of quantum secure computing has the potential to fundamentally change the assumptions underlying blockchain and crypto security design. It opens up new avenues for privacy, anonymity, and secure computation, but also presents challenges that must be overcome to realize its full potential in the blockchain and cryptocurrency landscape.
\subsection{Theoretical Framework}
Developing a theoretical framework for how digital currencies could operate in a post-quantum computing environment involves proposing a systematic exploration of how futuristic techniques might influence digital currencies. This framework envisions the coexistence of CBDCs and cryptocurrencies within a post-quantum computing landscape by leveraging new cryptographic measures. \\
One aspect of this framework focuses on integrating quantum-safe cryptographic algorithms, categorizing CBDCs and cryptocurrencies into quadrants based on the speed of algorithm adoption and governance models, with CBDCs potentially transitioning more rapidly than decentralized cryptocurrencies \cite{kiff2020survey}. Another dimension considers the application of techniques such as MPC, OT, and HE, with CBDCs and cryptocurrencies varying in their utilization of these methods, ranging from minimal use to comprehensive integration \cite{zhang2024central}. Additionally, the framework evaluates the level of privacy measures, ranging from non-private to fully anonymous identities for CBDCs and varying privacy features for cryptocurrencies, including transparent ledger amounts to encrypted coin amounts \cite{tinn2024theory,hull2024properties}.\\
This theoretical framework systematically models how diverse post-quantum computing innovations can facilitate the coexistence of CBDCs and cryptocurrencies, presenting a matrix of options depicting differing abilities to adopt measures across rapid, moderate, or elongated timelines. Policymakers could employ this model to strategize optimal technological pathways that support the continued usability, security, and privacy of digital currencies in the evolving landscape. Regular reassessment will be essential to refine the framework dimensions as technologies mature. By offering a constructive approach to conceptualize plausible technological trajectories and their implications, this proposed theoretical framework aims to stimulate insights for effectively navigating this dynamic and evolving digital currency space.

\section{Quantum Money and Connection to CBDC}
As mentioned above, quantum money, originally proposed by Stephen Weisner back in 70s of the previous century, offers a unique way to leverage the principles of quantum mechanics for the secure generation and verification of monetary notes or bills \cite{wiesner1983conjugate,shor2012quantum}.  
Each quantum banknote is composed of qubits prepared in a special quantum superposition state. According to the no-cloning theorem of quantum mechanics, these states are theoretically impossible to clone or counterfeit.
One of the key advantages of quantum money is that users can verify the authenticity of a bill by measuring its quantum state without disturbing it. This is made possible by the properties of quantum measurement. If a bill is counterfeited, it would collapse to eigenstates during the measurement process, revealing its fraudulent nature.
Subsequent schemes have built upon Weisner's original concept by encoding money states in multidimensional subspaces entangled across many particles. This approach enhances the security of quantum money by leveraging the complexities of entanglement.\\
The goal of quantum money is to achieve stronger security than what is possible with classical cash-like currencies by leveraging the power of quantum physics \cite{muthukrishnan2022exploration}. Counterfeiting quantum money would essentially require solving computationally difficult problems.
However, quantum money relies on unproven conjectures about the limitations of quantum computers and the effectiveness of proposed cryptographic schemes against future threats \cite{grimes2019cryptography}. It remains an active area of research with ongoing challenges, such as the development of long-term quantum memory, large-scale implementation, and the revocation of spent notes.
In the future, the insights from quantum information science could potentially influence the design of currencies, payments, and digital assets. If the technological hurdles associated with quantum money can be overcome, it holds great promise as a highly secure monetary instrument.\\
There are potential connections between quantum money and CBDC that could shape the future of monetary systems \cite{shafranova2024navigating}. One significant aspect is security. Quantum money schemes aim to leverage the principles of quantum physics to achieve unprecedented levels of security against counterfeiting \cite{ashfaq2023central}. If these schemes mature, the generation and verification of quantum states could enhance the security of a CBDC system, providing robust protection against fraudulent activities.
Anonymity is another area where quantum money concepts align with the objectives of CBDCs to some extent. Certain quantum coin schemes offer statistical anonymity similar to physical cash. This aspect of privacy aligns with the goals of CBDCs, although central banks still require oversight abilities to maintain regulatory control.
The conceptual model of individual banknotes or coins in quantum money schemes could influence the design of a digital currency beyond a centralized digital ledger of accounts \cite{allen2020design}. This means that the principles and ideas behind quantum money could shape the overall structure and presentation of a CBDC, potentially introducing innovative design elements.
The issuance mechanism of quantum money introduces new paradigms for how currency units are produced, distributed, and authenticated. These mechanisms could complement or enhance a CBDC issuance scheme run by the central bank, opening up possibilities for more efficient and secure methods of creating and distributing currency units. \\
Revocation is another area where quantum money poses challenges. The need to revoke or blacklist spent coins in quantum money links to the necessity for CBDC protocols that dynamically update based on spending while still preserving consumer privacy. Finding solutions to this issue could inform the development of CBDC systems.
The technology gap between the current state of quantum money and the implementation of CBDCs provides an opportunity for research and development. CBDCs can be introduced while allowing quantum money concepts to mature, addressing immediate concerns about phasing out physical cash. \\
Furthermore, the interfaces and standards developed in quantum cryptography and quantum money could contribute to the adoption and interoperability of CBDCs enabled by post-quantum technologies \cite{rationales20237}. These standards can ensure compatibility and facilitate the integration of quantum-resistant security measures into CBDC systems.
While quantum money concepts are still in their early stages, they play a role in influencing the design and properties that central bank digital currencies may adopt in the future, depending on the progress made in quantum technologies. The potential of quantum money offers valuable insights that can shape the development of CBDCs as quantum technologies continue to advance.

\section{Nonpermissive Banks Quantum and Attacks on Weisner's Scheme}
With the rapid progress of quantum computing, there is a growing concern that adversaries could eventually exploit its power to compromise currently secure cryptographic systems such as RSA and elliptic curve cryptography \cite{kurniawan2023quantum}. The implications for banks and financial institutions are significant. Encrypted banking records, passwords, financial transactions, and other sensitive data that were once considered secure could be exposed. Additionally, the integrity of digital signatures used to verify transactions could be compromised, potentially enabling fraudulent activities like double spending in cryptocurrencies \cite{darem2023cyber}.\\
To mitigate these risks, banks must proactively transition to post-quantum cryptographic algorithms that can withstand quantum attacks. Promising options include lattice-based and multivariate signature schemes, which offer resistance against quantum computers. The National Institute of Standards and Technology (NIST) is taking the lead in standardizing new public-key encryption and signature algorithms that are quantum-resistant \cite{alagic2022status}. Banks should closely monitor these efforts to stay informed about the latest developments. \\
In addition to algorithmic upgrades, banks need to upgrade their secure hardware, such as Hardware Security Modules (HSMs), to support quantum-safe algorithms as they become standardized \cite{sommerhalder2023hardware}. Furthermore, data requiring long-term storage, such as financial records, should undergo encryption upgrades or migration to ensure they remain secure even in the face of future quantum adversaries.
By adopting quantum-safe practices, banks not only prepare themselves for potential quantum threats but also build customer trust in the security of their data and transactions. It is crucial to stay ahead of the curve and be proactive in implementing these measures before quantum computers become a practical threat.
Furthermore, ongoing research is exploring the broader impact of quantum computing on banking beyond cryptography. This includes assessing optimization and artificial intelligence (AI) applications, which could revolutionize various aspects of banking operations and decision-making. \\
In summary, the rise of quantum computing poses serious challenges to the security of banks and financial institutions. Transitioning to post-quantum cryptographic algorithms, upgrading secure hardware, and addressing the long-term storage of sensitive data are crucial steps. By embracing quantum-safe practices, banks can mitigate future risks, protect customer information, and remain at the forefront of technological advancements in the financial sector.\\

\subsection{Attacks on Weisner's scheme}
Weisner's original quantum money scheme, although groundbreaking, was susceptible to various attacks that could compromise its security and enable counterfeiting. One such attack is cloning, where an adversary efficiently copies a quantum money state to generate counterfeit bills \cite{brodutch2014adaptive}. Weisner's scheme lacked protections against cloning, making it vulnerable in this regard.\\
Another attack is forging through hiding states, where an adversary hides a measured quantum money state without spending it and later presents it as a new bill. Weisner's scheme did not address this type of attack either \cite{aaronson2012quantum}.
Bribery of bank tellers poses another threat to Weisner's quantum money scheme. Adversaries could potentially bribe bank tellers responsible for verifying quantum money and convince them to accept counterfeit bills. The original scheme relied heavily on the honesty and integrity of the verification process, making it susceptible to such attacks.\\
State disturbance attacks represent another vulnerability. Adversaries could attempt to measure money states without spending them, causing slight disturbances while still maintaining their validity \cite{abdel2020entangling}. These slightly disturbed states could then be presented as genuine at a later time, bypassing the verification process.
Furthermore, the emergence of long-term quantum memory could pose a significant risk \cite{scholten2024assessing}. Adversaries could store copied money states until more advanced quantum computing technologies enable mass counterfeiting. The original scheme assumed quick verification, making it insufficient for addressing this potential threat.\\
In addition to these attacks, dishonest manufacturers of quantum bills could exploit the system. They might produce multiple copies of the same bill or retain production keys, allowing for mass counterfeiting at a later stage.
To mitigate these vulnerabilities, more recent quantum money schemes have been developed, focusing on mechanisms such as binding states to identities, trap states, and employing quantum cryptographic protocols between users and banks \cite{khodaiemehr2023navigating}. These advancements aim to address the identified attacks and provide enhanced security. However, it should be noted that Weisner's initial proposal was not proven secure against all quantum adversaries, highlighting the need for further research and development in the field of quantum money.

\section{Conclusions and Future Directions}
This paper examined questions around whether CBDC and blockchain-based cryptocurrencies could coexist in a post-quantum computing era. Through analyzing the implications of quantum algorithms and techniques like MPC and OT, several key findings were discussed.\\
Regulatory frameworks and international collaboration can play pivotal roles in mitigating risks stemming from quantum computing advancements in the context of digital currencies. \\
Coordinated disclosure efforts between regulators and industry stakeholders can establish responsible guidelines for disclosing quantum vulnerabilities, and allowing for thorough testing of solutions. Jurisdictions can synchronize transition roadmaps with clear upgrade deadlines to bolster preparedness and standardization across currency systems. \\
Cross-border research initiatives led by regulatory bodies can facilitate efficient development and assessment of post-quantum solutions tailored to diverse currency frameworks. International collaboration through standard-setting organizations can speed up the vetting and adoption of quantum-safe standards. \\
Regulators can play a key role in providing essential policy guidance on transition management, incentive structures, and maintaining trust during upgrades. They also oversee compliance monitoring to ensure that financial entities meet quantum-safety benchmarks by designated deadlines. Additionally, regulators collaborate on developing crisis contingency plans, implementing public education campaigns to raise user awareness, and conducting economic impact analyses to estimate vulnerability mitigation costs.\\
By aligning internationally, regulators have the potential to streamline efforts in managing quantum risks for digital currencies at a global level, leveraging their mandates to facilitate effective and timely transition management.\\
On the technical side, it appears CBDCs and cryptocurrencies may be able to adapt and incorporate measures like post-quantum cryptography to withstand quantum attacks on their core protocols. Technologies like blockchain, MPC and OT also show promise to be upgraded to quantum-secure variants that could still enable distributed, privacy-preserving applications.
Yet, significant challenges remain regarding transitioning legacy systems and achieving widespread integration of new quantum-resistant schemes. Open questions also persist around performance tradeoffs compared to classical assurances. Further theoretical and experimental research is still needed to fully evaluate real-world viability.\\
Moving forward, central banks and blockchain teams should actively pursue strategies to retrofit or future-proof their designs. Hybrid public-private ledger models combining the strengths of different approaches merit investigation. Standards bodies could aid dissemination of quantum-secure best practices.
Regulators must balance support for innovation with protections against quantum risks. Facilitating cooperation between sectors may help reconcile technical and policy developments. International collaboration will likewise be important as quantum technologies become increasingly global in nature.\\
While long-term coexistence seems plausible if challenges are addressed, the field remains in flux. Continued monitoring and theoretical work testing integration into live systems over time will be key to validate viability. Overall, a prudent yet optimistic approach is warranted, given the unprecedented opportunities - and uncertainties - that quantum computing may eventually usher in for financial technologies and beyond. Further advances on both the research and policy fronts will play a crucial role in shaping outcomes.


\bibliographystyle{IEEEtran}
\bibliography{ref.bib}

\end{document}